# High-efficiency WSi superconducting nanowire single-photon detectors operating at 2.5 K

V. B. Verma[1], B. Korzh[2], F. Bussières[2], R. D. Horansky[1], A. E. Lita[1], F. Marsili[3], M. D. Shaw[3], H. Zbinden[2], R. P. Mirin[1], and S. W. Nam[1]

[1]*National Institute of Standards and Technology, 325 Broadway, Boulder, CO 80305, USA*

[2]*Group of Applied Physics, University of Geneva, CH-1211 Geneva 4, Switzerland*

[3]*Jet Propulsion Laboratory, California Institute of Technology, 4800 Oak Grove Dr., Pasadena, California 91109, USA*

We investigate the operation of WSi superconducting nanowire single-photon detectors (SNSPDs) at 2.5 K, a temperature which is ~ 70 % of the superconducting transition temperature ($T_C$) of 3.4 K. We demonstrate saturation of the system detection efficiency at 78 ± 2 % with a jitter of 191 ps. We find that the jitter at 2.5 K is limited by the noise of the readout, and can be improved through the use of cryogenic amplifiers. Operation of SNSPDs with high efficiency at temperatures very close to $T_C$ appears to be a unique property of amorphous WSi.

Superconducting nanowire single-photon detectors (SNSPDs) are excellent detectors in the near-IR due to their high efficiencies (> 90%),[1] fast recovery times (< 100 ns),[2,3] low jitter (< 100 ps),[4] and low intrinsic dark count rates (< 1 count per second, cps).[1] Recently, the system detection efficiencies (SDE) achieved with NbN and NbTiN SNSPDs has improved dramatically, reaching values above 70 %.[5,6] However, the highest efficiencies to date (> 90 %) have been obtained through the use of amorphous WSi as the superconducting material.[1,7,8] The sources of the high efficiencies obtained with WSi are thought to be the small superconducting gap energy and lower carrier density compared to NbN, which causes a larger number of quasiparticles to be created per absorbed photon and the creation of a larger "hotspot". This increases the internal efficiency of the detector, which is the probability that an absorbed photon triggers a response pulse. In addition, the amorphous nature of WSi allows it to be deposited on virtually any substrate without significant degradation in material properties. Thus, WSi SNSPDs can easily be embedded inside of a dielectric optical stack to enhance absorption at a particular wavelength.

While high system detection efficiency is an attractive quality of WSi SNSPDs, obtaining the highest efficiency and lowest jitter generally requires cooling the detectors to 1 K or below. Unfortunately, expensive and relatively complex cryogenics are required to reach this temperature. In previous work we have shown that high efficiency may still be obtained at temperatures up to 2 K,[1] but at this temperature the switching current was only 2 µA, limiting the jitter to ~ 500 ps, which is not adequate for many experimental applications. Ideally, one would like to operate WSi SNSPDs with high efficiencies and low jitter (< 200 ps) in a closed cycle cryocooler operating at 2.5 K. These systems are commercially available, are simple to operate, and require only the installation of coaxial cables and optical fiber for operating the SNSPDs.

In this work, we show that operation of WSi SNSPDs at 2.5 K is possible without sacrificing efficiency. We demonstrate a saturation of the system detection efficiency at 78 ± 2 %. Furthermore, using a cryogenic preamplifier combined with a room temperature amplifier, we demonstrate a jitter of 191 ps full width at half maximum (FWHM). Remarkably, this device performance was obtained at a temperature that is ~ 70 % of the superconducting transition temperature (2.5 K operation with a $T_C$ of 3.4 K). This is in contrast to NbN SNSPDs that show significant degradation in efficiency when operated so close to $T_C$.[9]

The $W_{0.8}Si_{0.2}$ detectors were optimized for maximum absorption at a wavelength of 1340 nm in order to demonstrate quantum teleportation from a telecom-wavelength photon into a solid state quantum memory.[10] The fabrication process begins with the deposition of gold mirrors on a 3" Si wafer by electron beam evaporation and liftoff and consists of 80 nm of gold with a 2 nm Ti adhesion layer below and above the mirror. A quarter-wave spacer layer consisting of 195 nm of $SiO_2$ was then deposited by plasma-enhanced chemical vapor deposition (PECVD), and Ti/Au contact pads were patterned by liftoff. The 4.6 nm-thick WSi was deposited by DC magnetron cosputtering from separate W and Si targets at room temperature, and capped with 2 nm of amorphous Si to prevent oxidation. After etching a 20 μm-wide strip between Au contact pads, electron beam lithography and etching in an $SF_6$ plasma were used to define nanowire meanders with a width of 130 nm and pitch of 260 nm. An antireflection coating was deposited on the top surface consisting of 225 nm $SiO_2$, 179 nm $SiN_x$, 231 nm $SiO_2$, and 179 nm $SiN_x$. A keyhole shape was etched through the Si wafer around each SNSPD, which could then be removed from the wafer and self-aligned to a single-mode optical fiber to within ± 3 μm.[11] The size of the SNSPD is 16 μm × 16 μm, larger than the 10 μm mode field diameter of a standard single-mode fiber, to allow for slight misalignment.

The SNSPD was characterized at a temperature of 2.5 ± .005 K in an air-cooled Gifford-McMahon cryocooler with 100 mW of cooling power. The temperature was measured with a calibrated Cernox temperature sensor. A 1310 nm CW laser with a mean photon number of ~ 100,000 photons/s was used for the measurements. The calibration of the input power to the SNSPD was performed as outlined in Ref. 1 using a calibrated InGaAs power meter. Fig. 1 shows the system detection efficiency (SDE) and background count rate (BCR) as a function of bias current ($I_B$) normalized to the switching current ($I_{SW}$, the bias current at which the device switches from the superconducting to the normal state). The SDE was measured after maximizing the detector counts as a function of the polarization of the incident light. The maximum SDE is 78 ± 2 % with a BCR of 530 cps at $I_B \sim 0.9 I_{SW}$. Note that despite the relatively high operating temperature, the SDE reaches saturation at the highest bias currents. For the same value of $I_B$, the SDE reaches a minimum value of 66 % when the polarization of the incident light is adjusted to minimize the count rate. Significantly lower SDCR could be obtained (~ 3 cps) at a bias current $I_B \sim 0.8 I_{SW}$, with an SDE of ~ 70 %.

The dark count rate was measured with the fiber removed from the detector, which we call the intrinsic device dark count rate (DCR), as well as with a fiber coupled to the detector which we call the background count rate (BCR).[1, 7, 8] The BCR was also measured with a fiber that had been coiled (8 turns with a 27 mm diameter) just before the SNSPD. Coiling the fiber increases loss significantly at wavelengths greater than 2 µm, effectively forming a blackbody filter. Loss at 1310 nm due to the coil was estimated to be less than 0.01 dB. As shown in Fig. 2, the BCR is larger than the DCR, which we attribute to the detection of blackbody photons coupled into the optical fiber. The process of coiling the fiber reduces the BCR to the DCR within the uncertainty of the measurement, showing that it is effective at reducing the number of blackbody photons coupled to the detector.

Fig. 3 shows a voltage pulse from the SNSPD obtained with a cryogenic preamplifier at 40 K with 28 dB of gain, followed by a room temperature amplifier with 33 dB of gain, at a bias current of 2.2 µA. The jitter of the detector system ($J_S$) was measured using a 1550 nm pulsed laser with a FWHM of ~ 33 ps and a time-correlated single-photon counting card to register the inter-arrival time between the laser pulse and SNSPD pulse. Figure 4 shows the instrument response function (IRF) of the detector system biased at $I_B = 0.9 I_{SW}$, which is the bias current corresponding to maximum SDE. We estimated the system jitter to be the full width at half maximum (FWHM) of the IRF, $J_s = 191$ ps.

The system jitter presented in this Letter is not dominated by the intrinsic jitter of the detector, but by the electrical noise of the electronic readout circuitry. As discussed in Ref. 1, the total system jitter can be expressed as $J_s = (J_d^2 + J_n^2)^{1/2}$, where $J_d$ is the intrinsic device jitter and $J_n$ is the noise contribution to the jitter. The electrical noise contribution to the jitter can be estimated as $J_n = e_n/S$ where $e_n$ is the FWHM of the electrical noise and $S$ is the slope of the detection signal. To verify that the noise contribution to the jitter is indeed the main contribution, we directly measured the values $e_n$ and $S$, and calculated $J_n$ using a 6 GHz oscilloscope. Figure 5 shows the comparison between $J_s$ measured using the TCSPC method outlined earlier and $J_n$ as a function of normalized bias current. Taking a line of best fit through this data allows us to make an estimate of the intrinsic device jitter, $J_d = (J_s^2 - J_n^2)^{1/2}$ which is roughly constant and less than 100 ps for the whole bias current range. This highlights the importance of using a cryogenic pre-amplifier, in order to minimize the electrical noise. Indeed, using only room-temperature amplifiers increases the system jitter to around 380 ps at $I_B = 0.95 I_{SW}$. Further improvements could be gained by placing the preamplifier at 2.5 K, which has been successfully demonstrated recently.[12] The added benefit of the preamplifier being next to the SNSPD is that its input impedance can be easily increased, which provides further improvement to the signal to noise ratio and a reduction of the detector reset time.[12] We estimate that using a two stage amplifier chain at 2.5 K and 40 K would reduce the noise contribution to the jitter down to around 50 ps with the same detector, which would become comparable to the estimated intrinsic jitter.

In conclusion, we have demonstrated that WSi SNSPDs can be operated at a temperature that is nearly 70 % of the $T_C$ of the material without significant reduction in system detection efficiency. The ability to operate with high efficiency at temperatures approaching $T_C$ appears to be a unique property of amorphous WSi. The maximum measured SDE was 78 ± 2 %, which is comparable to the best results obtained with NbN and NbTiN SNSPDs at similar temperatures.[5, 6] Although the switching current is reduced to 2.2 μA at 2.5 K, a jitter of 191 ps was nevertheless obtained using a combination of a cryogenic preamplifier at 40 K and a room temperature amplifier. We have shown that the jitter is dominated by the electrical noise of the read-out circuit, not the intrinsic device jitter, and we have discussed how it could be reduced further by the use of a preamplifier at 2.5 K.

We acknowledge Claudio Barreiro and Mikael Afzelius for useful discussions, and the Swiss Federal Institute of Metrology (METAS) for the calibration of the power meter.

**Figures**

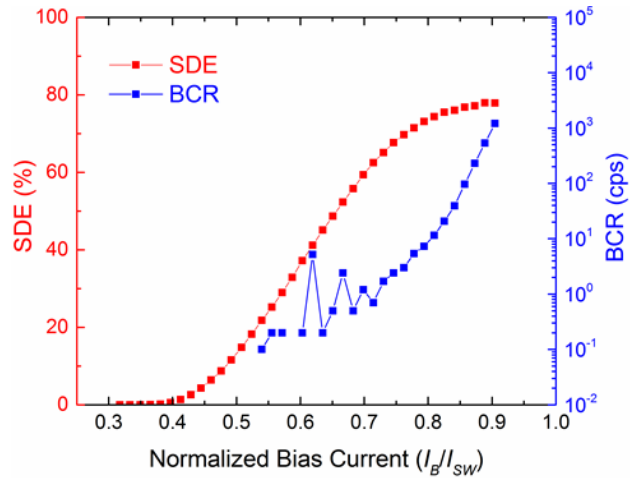

Fig. 1 System detection efficiency (SDE) and background count rate (BCR) as a function of normalized bias current. The switching current of the detector ($I_{SW}$) was 2.2 μA.

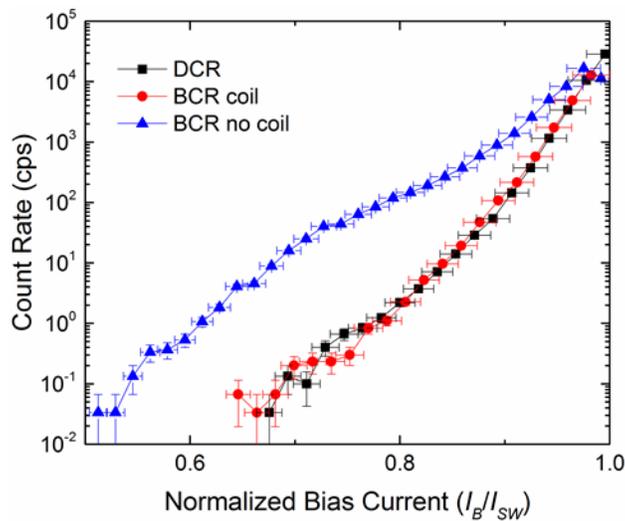

Fig. 2 Dark count rate measured without a fiber coupled to the detector (i.e. the intrinsic detector dark count rate (DCR)), with a fiber coupled to the detector without being coiled (background

count rate (BCR) no coil), and with a fiber coupled to the detector with a coil (background count rate (BCR) coil).

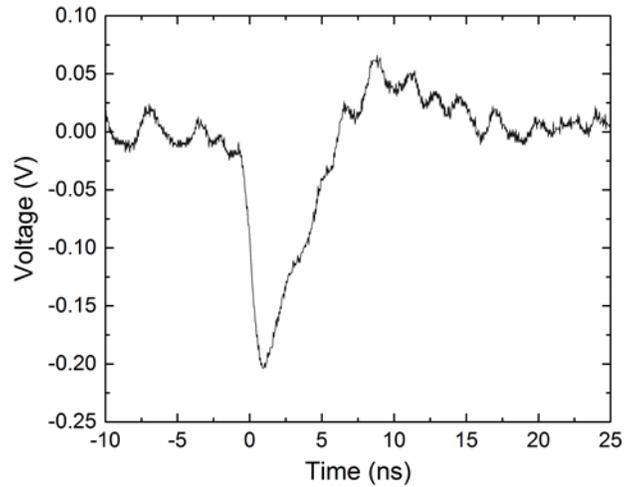

Fig. 3 Voltage pulse from the SNSPD obtained using a cryogenic preamplifier and room-temperature amplifier.

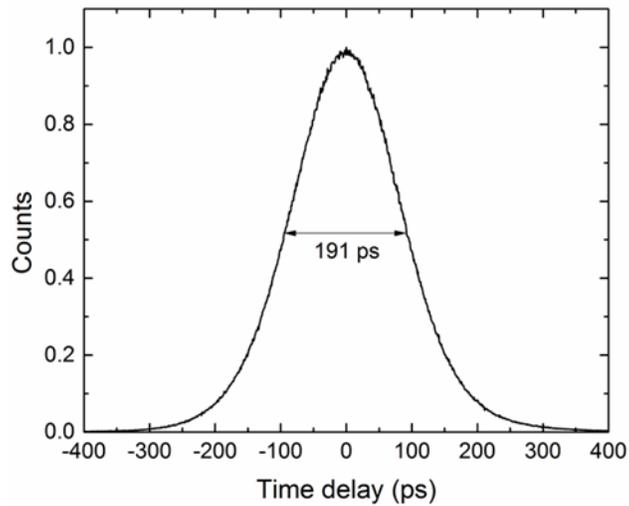

Fig. 4 Instrument response function (IRF) of the detector system. The jitter is 191 ps FWHM.

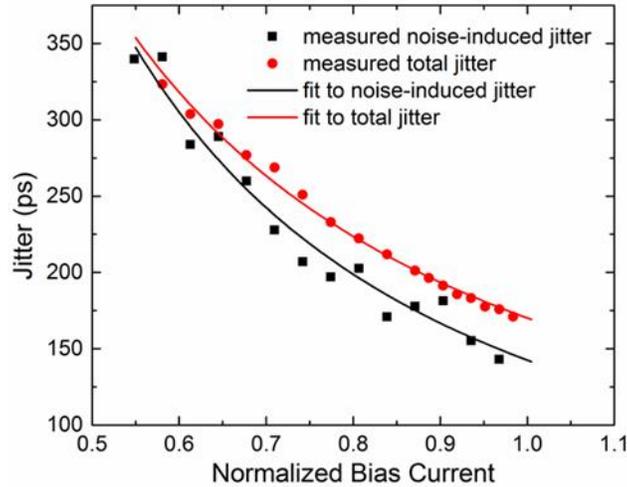

Fig. 5 Experimentally measured system jitter as a function of normalized bias current ($J_s$, red circles) and the noise contribution to the jitter ($J_n$, black squares). Solid lines are polynomial fits to the measured data.